\documentstyle[12pt]{article} 
\newcommand{\sn}{{\rm sn}}
\newcommand{\cn}{{\rm cn}}
\newcommand{\dn}{{\rm dn}}

\begin{document} 

\vspace{30mm}  

\centerline{\Large \bf The Elliptic Function in Statistical Integrable Models}

\vspace{15mm}


\centerline{\large Kazuyasu Shigemoto\footnote{E-mail address:
shigemot@tezukayama-u.ac.jp}}

\centerline {{
\large Tezukayama University, Tezukayama 7, Nara 631, Japan
}}

\vspace{20mm}

\centerline{\bf Abstract} 

\vspace{3mm} 

We examine the group theoretical reason why various 
two dimensional statistical integrable models, such as the Ising model, 
the chiral Potts model and the Belavin model, becomes integrable.
The symmetry of these integrable models is $SU(2)$ and the 
Boltzmann weight can be parametrized by the elliptic function in many cases. 
In this paper, we examine the connection between the $SU(2)$ symmetry 
and the elliptic function in the statistical integrable models.

\vspace{5mm}

\noindent
\begin{tabular}{|llll|} \hline 
Keywords: & elliptic function, & elliptic theta function,  & Ising model \\
 & chiral Potts model,  & Belavin model, & Heisenberg algebra\\
\hline
\end{tabular} 

\vspace{20mm}
\noindent {\bf Contents}\\
I Introduction\\
II The Ising model and the elliptic function\\
III The chiral Potts model and the elliptic function\\
IV The Belavin model and the elliptic theta function\\
V The Heisenberg algebra\\
VI Summary and discussion\\

\newpage
\setcounter{equation}{0}
\section{Introduction}

The two dimensional integrable statistical models are classified
into three types of model, the spin model, the vertex model and the 
face model\cite{Baxter}.

Typical $N$-state spin model is called the chiral Potts model\cite{chiral}, 
which includes the Ising model\cite{Onsager} as the special $N=2$ state model. 
The origin of the integrability 
of the non-linear model comes from the Lie group symmetry, which 
is some generalization of $SU(2)$, that is, the cyclic representation 
of $SU(2)$\cite{H-S}. In this model, we can parametrize the Boltzmann weight by
the elliptic function which has the difference property for only 
$N=2$ Ising case.

While typical $Z_N \times Z_N$ vertex model is called the Belavin 
model\cite{Belavin}, which
includes the Baxter's 8-vertex model\cite{Baxter1} as the special     
 $Z_2 \times Z_2$ vertex model. In this model, the origin of the 
integrability comes from the symmetry of the cyclic representation of 
$SU(2)$\cite{Tracy,Cherednik}.
In this model, we can parametrize 
the Boltzmann weight by the elliptic theta function with characteristics.

The typical face model(IRF model) is $A^{(1)}_{N-1}$ model\cite{Jimbo},
which is the generalization of Baxter's IRF model\cite{Baxter2}.  
But this face model is equivalent to the Belavin model by the vertex-face 
correspondence by using the intertwining vector\cite{Jimbo}.

Then the origin of the integrability of the two dimensional statistical model
comes from the cyclic $SU(2)$ symmetry. And we have the elliptic 
representation of the Boltzmann weight in many important cases.
Then we expect the correspondence between the cyclic $SU(2)$ symmetry 
and the elliptic function. In other words, we expect that the symmetry 
of the elliptic function is the cyclic $SU(2)$ symmetry\cite{Mumford}. 

In this paper, we examine various statistical integrable models
in the context of the correspondence of the cyclic $SU(2)$ 
symmetry and the elliptic function.

\vspace{10mm}

\setcounter{equation}{0}
\section{The Ising model and the elliptic function}
The star-triangle relation(the integrability condition) in the Ising model is 
written in the form
\begin{eqnarray}
  &&\sum_{d=\pm 1} \exp\{d(L_1 a+K_2 b +L_3 c) \}
  =R\exp(K_1bc +L_2 ca +K_3 ab )  . 
\label{e2-1}
\end{eqnarray}
This relation is written in the form
\begin{eqnarray}
  &&\exp( L^*_3\sigma_x ) \exp( K_2\sigma_z ) \exp( L^*_1\sigma_x )
   =\exp( K_1\sigma_z ) \exp(L^*_2\sigma_x ) \exp( K_3\sigma_z ) ,
\label{e2-2}
\end{eqnarray}
where we use $\tanh X^*=\exp(-2 X )$ , which gives 
$\sinh 2X \sinh 2X^*=1$.\\
If $\{L_i,K_i\}(i=1,2,3)$ satisfies the above integrable condition,
we have 
\begin{eqnarray}
  &&\exp(2L^*_3 J_x ) \exp( 2K_2 J_z ) \exp( 2L^*_1 J_x)
   =\exp( 2K_1 J_z ) \exp( 2L^*_2 J_x ) \exp( 2K_3 J_z )  ,
\label{e2-3}
\end{eqnarray}
for arbitrary spin of $SU(2)$ with the commutation relation
\begin{eqnarray}
  &&[J_x,J_y]=i J_z,\quad [J_y,J_z]=i J_x,\quad  [J_z,J_x]=i J_y  .
\label{e2-4}
\end{eqnarray}
If we define $J_{\pm}=J_x \pm i J_y$, the above commutation relation 
is written in the form
\begin{eqnarray}
  &&[J_z,J_{\pm}]=\pm J_{\pm},\quad [J_+,J_-]=2J_z .
\label{e2-5}
\end{eqnarray}

We will show that the integrability condition does not
depend on the magnitude of the spin of $SU(2)$ in the following way.
We denote $U$ and $V$  as the left-hand side and the
right-hand side of the integrability condition respectively, that is, 
\begin{eqnarray}
U=\exp( 2L^*_3 J_x ) \exp(2K_2 J_z ) \exp( 2L^*_1 J_x ),\ 
V=\exp( 2K_1 J_z ) \exp( 2L^*_2 J_x ) \exp( 2K_3 J_z )  .
\label{e2-6}
\end{eqnarray}
From the relation
\begin{eqnarray}
UJ_xU^{-1}=VJ_xV^{-1},\ UJ_yU^{-1}=VJ_yV^{-1}, \ UJ_zU^{-1}=VJ_zV^{-1} ,
\label{e2-7}
\end{eqnarray}
we have  
\begin{eqnarray}
  && \cosh 2K_k=\cosh 2K_i \cosh 2K_j +\sinh 2K_i \sinh 2K_j \cosh 2L^*_k ,
\label{e2-8}\\
  && \cosh 2L^*_k=\cosh 2L^*_i \cosh 2L^*_j 
  +\sinh 2L^*_i \sinh 2L^*_j \cosh 2K_k  ,
\label{e2-9}\\
  && \sinh 2K_k \cosh 2L^*_i=\cosh 2K_i \sinh 2K_j 
                           +\sinh 2K_i \cosh 2K_j \cosh 2L^*_k  ,
\label{e2-10}\\
  && \sinh 2L^*_k \cosh 2K_i=\cosh 2L^*_i \sinh 2L^*_j 
                           +\sinh 2L^*_i \cosh 2L^*_j \cosh 2K_k ,
\label{e2-11}\\
  && \sinh 2L^*_i \sinh 2L^*_j+\cosh 2L^*_i \cosh 2L^*_j \cosh 2K_k  
\nonumber\\
  &&=\sinh 2K_i \sinh 2K_j+\cosh 2K_i \cosh 2K_j \cosh 2L^*_k   ,
\label{e2-12}\\
  &&\frac{\sinh 2K_i}{\sinh 2L^*_i}=\frac{\sinh 2K_j}{\sinh 2L^*_j} ,
\label{e2-13}\\
  &&(k=2,\ i \ne j =1,3) ,
\nonumber
\end{eqnarray}
where we use only the commutation relation Eq.(\ref{e2-4}). 
Conversely, if Eq.(\ref{e2-7}) is satisfied, 
\begin{eqnarray}
  [V^{-1}U,J_x]=0, \ [V^{-1}U,J_y]=0, \ [V^{-1}U,J_z]=0  ,  
\label{e2-14}
\end{eqnarray}
which gives $V^{-1}U={ \rm const.} 1 $ by Shur's lemma. 
If we consider the special 
case $K_1=0$, $K_2=0$, $K_3=0$, $L^*_1=0$, $L^*_2=0$, $L^*_3=0$,
the proportional constant in the right of $V^{-1} U$ becomes $1$, 
and we have the integrability condition $U=V$, which is independent 
of the magnitude of the spin.\\
We parametrize the Boltzmann weight of the Ising model with the 
elliptic function. In that parametrization,
we put the following ansatz of the symmetry,\\
\noindent
i) $K_i \Leftrightarrow L_i$ corresponds to $u \Leftrightarrow K-u$ of the 
argument of the elliptic function.\\
ii) $K_i \Leftrightarrow K^*_i$ corresponds to $\sn(u) \Leftrightarrow \cn(u)$\\
iii)$L_i \Leftrightarrow L^*_i$ corresponds to $\sn(K-u) \Leftrightarrow \cn(K-u)$\\
We use Eq.(\ref{e2-13}) as the starting point.
Then we parametrize 
\begin{eqnarray}
&&\sinh 2K_i=F(\sn(u_i),\cn(u_i)), 
\nonumber\\
&&\sinh 2K^*_i=\frac{1}{\sinh 2K_i}=F(\cn(u_i),\sn(u_i))
=\frac{1}{F(\sn(u_i),\cn(u_i))} , 
\nonumber\\
&& \sinh 2L_i=F(\sn(K-u_i),\cn(K-u_i)),
\nonumber\\
&&\sinh 2K_i \sinh 2L_i=F(\sn(u_i),\cn(u_i))F(\sn(K-u_i),\cn(K-u_i))
=(i-{\rm independent}) .
\nonumber
\end{eqnarray}
From the relation $\sinh 2K_i \sinh 2K^{*}_i=F(\sn(u_i),\cn(u_i))F(\cn(u_i),\sn(u_i))=1$, 
we take 
\begin{eqnarray}
  \sinh 2K_i=F(\sn(u_i),\cn(u_i))=\frac{\sn(u_i)}{\cn(u_i)} .
\label{e2-15}
\end{eqnarray}
(Another possibility is 
$\sinh 2K_i=F(\sn(u_i),\cn(u_i))=\cn(u_i)/\sn(u_i)$
 but we do not take this possibility here.)\\
In this representation, we have  
\begin{eqnarray}
  &&\frac{\sinh 2K_i}{\sinh 2L^*_i}=\sinh 2K_i \sinh 2L_i
  =\frac{\sn(u_i)}{\cn(u_i)}\frac{\sn(K-u_i)}{\cn(K-u_i)}
  =\frac{1}{k'}={\rm const.}  .
\label{e2-16}
\end{eqnarray}
Then we take 
\begin{eqnarray}
  &&\cosh 2K_i=\frac{1}{\cn(u_i)} , \quad
  \sinh 2K_i=\frac{\sn(u_i)}{\cn(u_i)} ,
\label{e2-17}\\
  &&\cosh 2L^*_i=\frac{1}{\sn(K-u_i)} , \quad 
  \sinh 2L^*_i=\frac{\cn(K-u_i)}{\sn(K-u_i)}, 
\label{e2-18}\\
  &&(i=1,2,3)  .
\nonumber
\end{eqnarray}
From Eq.(\ref{e2-8}) and Eq.(\ref{e2-9}), we have
\begin{eqnarray}
  &&(1-\sinh 2K_1 \sinh 2K_3 \sinh 2L^*_1 \sinh 2L^*_3) \cosh 2K_2 
\nonumber\\
  &&=\cosh 2K_1 \cosh 2K_3 +\sinh 2K_1 \sinh 2K_3 \cosh 2L^*_1 \cosh 2L^*_3  .
\label{e2-19}
\end{eqnarray}
Using the above parametrization Eq.(\ref{e2-17}) and Eq.(\ref{e2-18}), 
Eq.(\ref{e2-19}) is written in the form 
\begin{eqnarray}
  &&\cn(u_2)=\frac{\cn^2(u_1) \cn^2(u_3)-{k'}^2 \sn^2(u_1) \sn^2(u_3)}
  {\cn(u_1) \cn(u_3)+ \sn(u_1)\dn(u_1) \sn(u_3)\dn(u_3)}
\nonumber\\
  &&=\frac{\cn(u_1) \cn(u_3)- \sn(u_1)\dn(u_1) \sn(u_3)\dn(u_3)}
  {1-k^2 \sn^2(u_1) \sn^2(u_3)}  .
\label{e2-20}
\end{eqnarray}
By the addition theorem of the elliptic function, we obtain the relation 
among  $u_1,\ u_2,\ u_3$ in the form  
\begin{eqnarray}
  u_2=u_1+u_3  .
\label{e2-21}
\end{eqnarray}
Using Eq.(\ref{e2-17}), Eq.(\ref{e2-18}), and Eq.(\ref{e2-21}), 
we have checked that Eq.(\ref{e2-8})-Eq.(\ref{e2-13}) are really satisfied. 

\setcounter{equation}{0}
\section{The chiral Potts model and the elliptic function}
The chiral Potts model is the integrable $N$-state spin model.
The star-triangle relation(integrable condition) in the chiral Potts model
is given by
\begin{eqnarray}
  &&\sum^{N-1}_{d=0} \overline{W}_{qr}(b-d) W_{pr}(a-d) \overline{W}_{pq}(d-c)
  =R_{pqr} W_{pq}(a-b) \overline{W}_{pr}(b-c) W_{qr}(a-c),
\label{e3-1}\\
  && W_{pq}(k)=\prod_{l=1}^{k} 
  \left( \frac{d_p b_q - a_p c_q \omega^{l}}
         {b_p d_q - c_p a_q \omega^{l}}  \right), \quad
  \overline{W}_{pq}(k)=\prod_{l=1}^{k} 
  \left( \frac{\omega a_p d_q - d_p a_q \omega^{l}}
         {c_p b_q - b_p c_q \omega^{l}}  \right), 
\label{e3-2}\\
  &&a^N_p+k' b^N_p=k d^N_p, \quad k' a^N_p+b^N_p=k c^N_p  . 
\label{e3-3} 
\end{eqnarray}
This condition is rewritten into a nice form, 
which is expressed with the Lie group element 
in the form \cite{H-S}
\begin{eqnarray}
  &&T_{pq} S_{pr} T_{qr}=S_{qr} T_{pr} S_{pq},
\label{e3-4} \\
  &&T_{pq}=\sum_{k=1}^{N} \widetilde{W}_{pq}(k) Z^k, \quad 
    S_{pq}=\sum_{k=1}^{N} \overline{W}_{pq}(k) X^k, 
\label{e3-5} \\
  &&\widetilde{W}_{pq}(k)=\sum^{N-1}_{l=0}\omega^{kl}W_{pq}(l)
   = \prod_{l=1}^{k} 
  \left( \frac{b_p d_q - d_p b_q \omega^{l-1}}
         {c_p a_q - a_p c_q \omega^{l}}  \right),
\label{e3-6}\\
  &&\overline{W}_{pq}(k)=\prod_{l=1}^{k} 
  \left( \frac{a_p d_q \omega - d_p a_q \omega^{l}}
         {c_p b_q - b_p c_q \omega^{l}}  \right),
\label{e3-7}
\end{eqnarray}
where $Z$ and $X$ are elements of the cyclic representation of 
$SU(2)$, which satisfy $Z X=\omega X Z$, $(\omega=e^{2  \pi i/N})$.
In order to show Eq.(\ref{e3-4}), we have used the following 
relation
\begin{eqnarray}
  &&P(Z)(\alpha Z +\beta)X=(\gamma Z +\delta) X P(Z),
\label{e3-8} \\
  &&P(Z)=\sum_{k=1}^{N} p_k Z^k, \quad p_k=\prod_{l=1}^{k} 
  \left(\frac{\gamma \omega -\alpha \omega^{l}}
         {\beta \omega^{l} -\delta}  \right) ,
\label{e3-9}\\
 &&Q(X)(\alpha X +\beta)Z=(\gamma X +\delta) Z Q(X),
\label{e3-10} \\
  &&Q(X)=\sum_{k=1}^{N} q_k X^k, \quad q_k=\prod_{l=1}^{k} 
  \left( \frac{\gamma \omega^{l-1} -\alpha}
         {\beta -\delta \omega^{l}}  \right).
\label{e3-11}
\end{eqnarray}

\subsection{The Ising model}
The Ising model is the special $N=2$ case of the chiral Potts model. 
The parametrization, which satisfies Eq.(\ref{e3-3}) and has the difference 
property $W_{p,q}(n)=f_n(p-q)$, $\widetilde{W}_{p,q}(n)=g_n(p-q)$, is 
given by\cite{Perk}
\begin{eqnarray}
  && (a_p,b_p,c_p,d_p)
  =(\theta_{11}(p/2K),\theta_{10}(p/2K),\theta_{00}(p/2K),\theta_{01}(p/2K)) .
\label{e3-12} 
\end{eqnarray}
Using the relation,
\begin{eqnarray}
  \sn(u)=-\frac{1}{\sqrt{k}} \frac{\theta_{11}(u/2K)}{\theta_{01}(u/2K)} ,
\quad
  \cn(u)=\sqrt{\frac{k'}{k}} \frac{\theta_{10}(u/2K)}{\theta_{01}(u/2K)} ,
\quad
  \dn(u)=\sqrt{k'}\frac{\theta_{00}(u/2K)}{\theta_{01}(u/2K)}  ,
\label{e3-13} 
\end{eqnarray}
we obtain the Ising model from the $N=2$ chiral Potts model by using 
the addition theorem of the elliptic function in the form 
\begin{eqnarray}
  &&\overline{W}_{pq}(1)/\overline{W}_{pq}(0)=\frac{\sinh{L^*}}{\cosh{L^*}}
  =e^{-2L}
\nonumber\\
  &&=\frac{-a_p d_q +d_p a_q}{c_p b_q+b_p c_q}
  =\frac{\dn(p-q)-\cn(p-q)}{k' \sn(p-q)} ,
\label{e3-14}\\
  &&\widetilde{W}_{pq}(1)/\widetilde{W}_{pq}(0)=\frac{\sinh{K}}{\cosh{K}}
  =e^{-2K^*}
\nonumber\\
  &&=\frac{b_p d_q -d_p b_q}{c_p a_q+a_p c_q}
  =\frac{1-\cn(p-q)}{\sn(p-q)}.
\label{e3-15}
\end{eqnarray}

\setcounter{equation}{0}
\section{The Belavin model and the elliptic theta function}
The Belavin model is the integrable $Z_N \times Z_N$
vertex type model. 
The Yang-Baxter equation, which is the integrability condition 
in this vertex model, is given by
\begin{eqnarray}
  S_{12}(u_1-u_2) S_{13}(u_1-u_3) S_{23}(u_2-u_3)
  =S_{23}(u_2-u_3) S_{13}(u_1-u_3) S_{12}(u_1-u_2)  .
\label{e4-1}
\end{eqnarray}
The Boltzmann weight $S(u)$ is given by 
\begin{eqnarray}
  &&S(u)=\sum^{N-1}_{\alpha_1,\alpha_2=0} w_{\alpha_1, \alpha_2}(u) 
  I_{\alpha_1, \alpha_2} 
  \otimes I^{-1}_{\alpha_1, \alpha_2} , 
\label{e4-2}\\
  &&w_{\alpha_1, \alpha_2}(u)=\frac{1}{N} 
  \frac{\sigma_{\alpha_1, \alpha_2}(u+\eta/N)\sigma_{0,0}(\gamma \eta)}
  {\sigma_{\alpha_1, \alpha_2}(\eta/N)\sigma_{0,0}(u+\gamma \eta)} ,
\label{e4-3}\\
  &&\sigma_{\alpha_1, \alpha_2}(u)
  =\theta_{\frac{\alpha_2}{N}+\frac{1}{2}, \frac{\alpha_1}{N}
   +\frac{1}{2}}(u,\tau) ,
\label{e4-4}\\
  &&I_{\alpha_1, \alpha_2} =Z^{\alpha_1} X^{\alpha_2} ,
\label{e4-5}
\end{eqnarray}
where $Z$ and $X$ are elements of the cyclic representation of $SU(2)$ 
in the form
\begin{eqnarray}
  &&Z=\left(   \begin{array} {ccccc}  
  1 &  &  &   &  \\
    & \omega &  &  & \\
  &   & \omega  &  & \\
  &   &  & \cdots & \\
  &  &  &  & \omega^{N-1} \\
  \end{array} \right),  \quad
  X=\left(   \begin{array} {ccccc}  
  0 & 0 & \cdots  & 0  & 1 \\
  1  & 0 & \cdots & 0 & 0 \\
  0  & 1 & \cdots & 0 & 0 \\
  &   &  \cdots  &  &  \\
  0 & 0 & \cdots & 1 & 0 \\
  \end{array} \right),
\label{e4-6}\\
  &&ZX=\omega XZ, \quad \omega=\exp( 2 \pi i /N ), 
  \quad \omega^{N}=1.
\label{e4-7}
\end{eqnarray}
The index independent factor 
$\sigma_{0,0}(\gamma \eta)/\sigma_{0,0}(u+\gamma \eta)$ 
in $w_{\alpha_1. \alpha_2}(u)$ is trivially factor out in 
the integrability condition Eq.(\ref{e4-1}), 
but we put this factor in order that $S(u)$ satisfy the Zamolodchikov
algebra by choosing  $\gamma$ to be the appropriate 
value. The Zamolodchikov algebra, which is the fundamental relation of the 
integrability condition in the Belavin model, will be discussed  later.\\
The theta function with characteristics  $\theta_{r_1,r_2}(u,\tau)$ in 
the above is given by
\begin{eqnarray}
\theta_{r_1,r_2}(u,\tau)=\sum_{n\in Z} 
e^{ i \pi (n+r_1)^2 \tau + 2 \pi i (n+r_1)(u+r_2)}  .
\label{e4-8}
\end{eqnarray}
The $u$-independent but index dependent factor
$1/\sigma_{\alpha_1,\alpha_2}(\eta/N)$ is necessary 
to make $w_{\alpha_1 , \alpha_2}(u)$ to be periodic in the index $\alpha_2$.
This property comes from the relation
\begin{eqnarray}
&&\sigma_{\alpha_1, \alpha_2+N}(u+\eta/N)
=e^{2 \pi i \alpha_1/N}\sigma_{\alpha_1,\alpha_2}(u+\eta/N),
\nonumber\\
&&\sigma_{\alpha_1, \alpha_2+N}(\eta/N)
=e^{2 \pi i \alpha_1/N}\sigma_{\alpha_1,\alpha_2}(\eta/N),
\nonumber\\
&&\frac{\sigma_{\alpha_1, \alpha_2+N}(u+\eta/N)}
{\sigma_{\alpha_1, \alpha_2+N}(\eta/N)}
=\frac{\sigma_{\alpha_1, \alpha_2}(u+\eta/N)}
{\sigma_{\alpha_1, \alpha_2}(\eta/N)}  .
\nonumber
\end{eqnarray}
In the special $u=0$ case, $S_{12}(0)$ becomes the 
permutation operator \\
$\displaystyle{P_{12}=\frac{1}{N}\sum_{a,b}Z^a X^b \otimes X^{-b} Z^{-a}}$.
Then the Yang-Baxter equation is trivially satisfied in the special 
$u_1=u_2=u_3=0$ case.

\subsection{The cyclic and the ordinary spin representation in $SU(2)$}
The $N$-state cyclic representation of $SU(2)$ is given by 
\begin{eqnarray}
  &&(Z)_{a,b}=\delta_{a,b} \exp( 2\pi i a/N ), \quad
  (X)_{a,b}=\delta_{a,b+1} +\delta_{a,b+1-N},
\label{e4-9}\\
&&(a,b=0,\cdots,N-1) .
\nonumber
\end{eqnarray}
While the ordinary spin $J$ representation of $SU(2)$ is given by 
\begin{eqnarray}
  &&(J_z)_{a,b}=\delta_{a,b}(J-a),\quad 
  (J_+)_{a,b}=\delta_{a,b+1} ,\quad
  (J_-)_{a,b}=\delta_{a,b-1} ,\quad
\label{e4-10}\\
  &&(a,b=0,\cdots,N-1) , \quad (N=2J+1)  .
\nonumber
\end{eqnarray}
Then we have the relation between the $N$-state cyclic representation and 
the ordinary spin $J$ representation in the form
\begin{eqnarray}
  &&Z=\exp( 2\pi i (J-J_z)/N ), \quad X=J_{+} +J_{-}^{N-1}  .
\label{e4-11}
\end{eqnarray}
In the above relation, we can interprete that the element of the 
cyclic representation is somewhat the quantum group element
in a sense that $Z$ is the element of the Lie group but $X$ is the
sum of the element of Lie algebra.\\
In the special $N=2$ case, $Z, X$ become elements 
of Lie albegra $Z=\sigma_z, X=\sigma_x$.

\subsection{The Zamolodchikov algebra}
The fundamental integrability relation in the Belavin model is the following 
Zamolodchikov algebra\cite{Cherednik}
\begin{eqnarray}
  A(u_1)\otimes A(u_2+\eta)=S(u_1-u_2)A(u_1+\eta )\otimes A(u_2)  .
\label{e4-12}
\end{eqnarray}
We define $T(u,v)$ as
\begin{eqnarray}
  T(u,v)=S_{12}(u) S_{13}(u+v) S_{23}(v)-S_{23}(v) S_{13}(u+v) S_{12}(u) .
\label{e4-13}
\end{eqnarray}
If the Zamolodchikov algebra Eq.(\ref{e4-12}) is satisfied, we have 
\begin{eqnarray}
  T(u,v) A(w+u-\eta) \otimes A(w) \otimes A(w-v+\eta)=0  .
\label{e4-14}
\end{eqnarray}
Then we have the Yang-Baxter relation $T(u,v)=0$, if $A(u)$ satisfies 
the Zamolodchikov algebra and $A(u)$ is the basis of the complete set.\\
We will construct this $A(u)$ by the elliptic theta function,
and we will show that $A(u)$ is the basis of the complete basis.
In order to construct $A(u)$ by the elliptic theta function, 
we examine the quasi-periodic property of the Boltzmann weight
$S(u)$. If both side of the Zamolodchikov algebra has
the same quasi-periodicity, it suggests that the left-hand 
side is equal to the right-hand side up to the constant factor.\\
Using the property
\begin{eqnarray}
  \theta_{r_1, r_2}(u+\xi_1 +\xi_2 \tau,\tau)
  =e^{-i \pi \xi_2(\xi_2 \tau+2u)} e^{2 \pi i (r_1 \xi_1-r_2 \xi_2)}
  \theta_{r_1, r_2}(u,\tau), \quad (\xi_1, \xi_2 \in  Z)  .
\label{e4-15}
\end{eqnarray}
We have the transformation of the Boltzmann weight
\begin{eqnarray}
&&w_{\alpha_1, \alpha_2}(u+\xi_1 +\xi_2 \tau)=
\frac{\theta_{\frac{\alpha_2}{N}+\frac{1}{2},\frac{\alpha_1}{N}+\frac{1}{2}}
(u+\xi_1 +\xi_2 \tau+\eta/N,\tau)}
{\theta_{\frac{\alpha_2}{N}+\frac{1}{2},\frac{\alpha_1}{N}+\frac{1}{2}}
(u+\eta/N,\tau)}
\nonumber\\
&& \times \frac{\theta_{\frac{1}{2},\frac{1}{2}}(u+\gamma \eta,\tau)}
 {\theta_{\frac{1}{2},\frac{1}{2}}(u+\xi_1 +\xi_2 \tau+\gamma \eta, \tau)} 
 w_{\alpha_1, \alpha_2}(u)
\nonumber\\
  &&= e^{2 \pi i \xi_2 \eta(\gamma-1/N)} 
  e^{2 \pi i (\xi_1 \alpha_2 -\xi_2 \alpha_1)/N}
  w_{\alpha_1, \alpha_2}(u)
  = e^{2 \pi i \xi_2 \eta(\gamma-1/N)} 
  \omega^{<{\bf \xi},{\bf \alpha}>} w_{\alpha_1, \alpha_2}(u)  ,
\label{e4-16}
\end{eqnarray}
where we use the notation 
$<{\bf \xi},{\bf \alpha}>=\xi_1 \alpha_2 -\xi_2 \alpha_1$.
Because of the index independent but $u$ dependent factor 
$\sigma_{0,0}(\gamma \eta)/\sigma_{0,0}(u+\gamma \eta)$, which is 
trivially factor out 
in the Yang-Baxter equation Eq.(\ref{e4-1}), the multiplied factor 
in the right-hand side of Eq.(\ref{e4-16}) becomes $u$ independent,
which is necessary to satisfy the Zamolodchikov algebra.
Therefore the Boltzmann weight transforms into the form
\begin{eqnarray}
  &&S(u+\xi_1+\xi_2 \tau)
  =\sum^{N-1}_{\alpha_1,\alpha_2=0}w_{\alpha_1, \alpha_2}(u+\xi_1+\xi_2 \tau)
  Z^{\alpha_1}X^{\alpha_2}\otimes X^{-\alpha_2}Z^{-\alpha_1}
\nonumber\\
  &&=e^{2 \pi i \xi_2 \eta(\gamma-1/N)} 
  \sum^{N-1}_{\alpha_1,\alpha_2=0}w_{\alpha_1, \alpha_2}(u)
  \omega^{<{\bf \xi},{\bf \alpha}>}Z^{\alpha_1}X^{\alpha_2}
  \otimes X^{-\alpha_2}Z^{-\alpha_1} ,
  \nonumber
\end{eqnarray}
Then we have
\begin{eqnarray}
  &&e^{-2 \pi i \xi_2 \eta(\gamma-1/N)} S(u+\xi_1+\xi_2 \tau)
\nonumber\\
  &&=(I_{{\bf \xi}}\otimes 1 ) S(u) (I^{-1}_{{\bf \xi}}\otimes 1 ) 
  =(1 \otimes I^{-1}_{{\bf \xi}}) S(u) (1 \otimes I_{{\bf \xi}}) .
\label{e4-17}
\end{eqnarray}

\subsection{Construction of the state vector $A(u)$} 
Next we construct the state vector $A(u)$ by the elliptic theta 
function.
For this purpose we consider the transformation of
$\theta_{\frac{a}{N}, Bk}(u+\xi_1+\xi_2 \tau; C \tau)$, where $B, C$ are
constant.
\begin{eqnarray}
  \theta_{\frac{a}{N}, Bk}(u+\xi_1 +\xi_2 \tau;C \tau)
  =\omega^{a \xi_1} e^{-i \pi \tau \xi_2^2/C-2 \pi i \xi_2 (u+B k)/C}
  \theta_{\frac{a}{N}+\frac{\xi_2}{C}, B k}(u ;C \tau) . 
\label{e4-18}
\end{eqnarray} 
In order that  
$\theta_{\frac{a}{N}, Bk}(u;C \tau), \quad (a, k \in Z_N)$
is closed under the transformation, and also the 
prefactor $\omega^{a \xi_1} e^{-i \pi \tau \xi_2^2/C-2 \pi i \xi_2 (u+B k)/C}$
becomes $k$ independent, we have two possibilities
\begin{eqnarray}
  &&{\rm case\ 1)}: B=C=N/(N-1) 
\nonumber\\
  &&\theta_{\frac{a}{N}, B k}(u; C \tau)
  =\theta_{\frac{a}{N}, \frac{N k}{N-1}}(u;  N\tau/(N-1))  ,
\label{e4-19}\\
  &&\theta_{\frac{a}{N}, \frac{N k}{N-1}}(u+\xi_1+\xi_2 \tau; N\tau/(N-1))
\nonumber\\
  &&=e^{-i \pi \tau \xi_2^2 (N-1)/N-2 \pi i \xi_2 u (N-1)/N}
  \omega^{a \xi_1} \theta_{\frac{a-\xi_2}{N}, \frac{N k}{N-1}}(u;N\tau/(N-1)) .
\label{e4-20}\\
  &&{\rm case\ 2)}: B=C=N 
\nonumber\\
  &&\theta_{\frac{a}{N}, B k}(u; C \tau)
  =\theta_{\frac{a}{N}, Nk}(u; N \tau)=\theta_{\frac{a}{N},0}(u; N \tau)
   =(k-{\rm independent}) ,
\label{e4-21}\\
  &&\theta_{\frac{a}{N}, 0}(u+\xi_1+\xi_2 \tau; N \tau)
  =e^{-i \pi \tau \xi_2^2/N-2 \pi i \xi_2 u/N}
\omega^{a \xi_1} \theta_{\frac{a+\xi_2}{N}, 0}(u; N \tau)   .
\label{e4-22}
\end{eqnarray}
Then we define $A_k(u)$ in the case 1) in the form,
\begin{eqnarray}
  && A_k(u)
  =\left(   \begin{array} {c}  
  \theta_{0, \frac{kN}{N-1}}(u; N \tau/(N-1)) \\ 
  \theta_{\frac{1}{N}, \frac{kN}{N-1}} (u; N \tau/(N-1))\\ 
  \theta_{\frac{2}{N}, \frac{kN}{N-1}}(u; N \tau/(N-1))\\
  \cdots \\ 
  \theta_{\frac{N-1}{N}, \frac{kN}{N-1}}(u; N \tau/(N-1))  \end{array} \right) ,
  \label{e4-23}\\
  &&A_k(u+\xi_1+\xi_2 \tau)=
  e^{-i \pi \tau \xi_2^2 \frac{N-1}{N}-2 \pi i \xi_2 u \frac{N-1}{N}}
  Z^{\xi_1} X^{\xi_2} A_k(u), \quad (\xi_1, \xi_2\in Z).
\label{e4-24}
\end{eqnarray} 

\subsection{The transformation of the Zamolodchikov algebra}
We examine the transformation 
$u_1\rightarrow u_1+\xi_1+\xi_2 \tau, \ (\xi_1,\xi_2\in Z)$
and $u_2\rightarrow u_2+\zeta_1+\zeta_2 \tau, \ (\zeta_1,\zeta_2\in Z)$ 
of the Zamolodchikov algebra in the  case 1)
\begin{eqnarray}
  A(u_1)\otimes A(u_2+\eta)=S(u_1-u_2)A(u_1+\eta )\otimes A(u_2)  .
\label{e4-25}
\end{eqnarray}
Under the transformation
$u_1\rightarrow u_1+\xi_1+\xi_2 \tau, \ (\xi_1,\xi_2\in Z)$, the Zamolodchikov 
algebra tranforms into the form 
\begin{eqnarray}
  &&({ \rm left-hand\ side})=A(u_1+\xi_1+\xi_2 \tau)\otimes A(u_2+\eta)
\nonumber\\
  &&=e^{-i \pi \tau \xi_2^2 \frac{N-1}{N}-2 \pi i \xi_2 u_1 \frac{N-1}{N}}
  (I_{{\bf \xi}} \otimes 1 )A(u_1)\otimes A(u_2+\eta) ,
\label{e4-26}\\
  && \nonumber\\
  &&({ \rm right-hand\ side})
  =S(u_1+u_1+\xi_1+\xi_2 \tau-u_2)
  A(u_1+\xi_1+\xi_2 \tau+\eta )
  \otimes A(u_2)
\nonumber\\
  &&=e^{2 \pi i \xi_2 \eta (\gamma-1/N)} e^{-i \pi \tau \xi_2^2 \frac{N-1}{N}
  -2 \pi i \xi_2 (u_1+\eta) \frac{N-1}{N}}
\nonumber\\
  && \times (I_{{\bf \xi}} \otimes 1 )S(u_1-u_2)(I^{-1}_{{\bf \xi}} \otimes 1 )
  (I_{{\bf \xi}} \otimes 1 )A(u_1+\eta)\otimes A(u_2)
\nonumber\\
  &&=e^{-i \pi \tau \xi_2^2 \frac{N-1}{N}-2 \pi i \xi_2 u_1 \frac{N-1}{N}}
  e^{ 2 \pi i \xi_2 \eta (\gamma-1)}
  (I_{{\bf \xi}} \otimes 1 )S(u_1-u_2)A(u_1+\eta)\otimes A(u_2)  .
\label{e4-27}
\end{eqnarray}
In order that the transformation is the same in the left-hand and 
the right-hand side of the Zamolodchikov algebra, we have $\gamma=1$
in case 1).\\
By the similar calculation, we have shown that the Zamolodchikov 
algebra has the same form  under the transformation 
$u_2\rightarrow u_2+\zeta_1+\zeta_2 \tau, \ (\zeta_1,\zeta_2\in Z)$. 

\subsection{The possibility of another Zamolodchikov algebra}
Using the vector of case 2), we have the possibility of another 
Zamolodchikov algebra.
In the case 2), we denote the state vector as $\widetilde{A}_k(u)$, 
which is given by 
\begin{eqnarray}
  && \widetilde{A}_0(u)
  =\left(   \begin{array} {c}  
  \theta_{0, 0} (u; N \tau)\\ 
  \theta_{\frac{1}{N}, 0} (u; N \tau)\\ 
  \theta_{\frac{2}{N}, 0} (u; N \tau)\\
  \cdots \\ 
  \theta_{\frac{N-1}{N}, 0} (u; N \tau) \end{array} \right) ,
\label{e4-28}\\
  &&\widetilde{A}_0(u+\xi_1+\xi_2 \tau)=
  e^{-i \pi \tau \xi_2^2/ N-2 \pi i \xi_2 u/ N}
  Z^{\xi_1} X^{-\xi_2} \widetilde{A}_0(u), \quad (\xi_1, \xi_2 \in Z).
\label{e4-29}
\end{eqnarray} 
Correspondingly, we define the another Boltzmann weight $\widetilde{S}(u)$
in the form
\begin{eqnarray}
  &&\tilde{S}(u)=\sum^{N-1}_{\alpha_1,\alpha_2=0} 
  \widetilde{w}_{\alpha_1, \alpha_2}(u) 
  J^{-1}_{\alpha_1, \alpha_2} 
  \otimes J_{\alpha_1, \alpha_2}  ,
\label{e4-30}\\
  &&\widetilde{w}_{\alpha_1, \alpha_2}(u)=\frac{1}{N} 
  \frac{\sigma_{\alpha_1, \alpha_2}(u+\eta/N)\sigma_{0,0}(\gamma \eta)}
  {\sigma_{\alpha_1, \alpha_2}(\eta/N)\sigma_{0,0}(u+\gamma \eta)} ,
\label{e4-31}\\
  &&J_{\alpha_1, \alpha_2} =Z^{\alpha_1} X^{-\alpha_2}  .
\label{e4-32}
\end{eqnarray}
We consider another Zamolodchikov algebra
\begin{eqnarray}
\widetilde{A}(u_1)\otimes \widetilde{A}(u_2+\eta)
=\widetilde{S}(u_1-u_2) \widetilde{A}(u_1+\eta )\otimes 
\widetilde{A}(u_2)  .
\label{e4-33}
\end{eqnarray}
Under the transformation
$u_1\rightarrow u_1+\xi_1+\xi_2 \tau, \ (\xi_1,\xi_2\in Z)$, another Zamolodchikov
algebra transforms in the form
\begin{eqnarray}
  &&({ \rm left-hand\ side})=\tilde{A}(u_1+\xi_1+\xi_2 \tau)
  \otimes \tilde{A}(u_2+\eta)
\nonumber\\
  &&=e^{-i \pi \tau \xi_2^2/N-2 \pi i \xi_2 u_1/N}
  (J_{{\bf \xi}} \otimes 1 ) \widetilde{A}(u_1)
  \otimes \tilde{A}(u_2+\eta) ,
\label{e4-34}\\
  &&({ \rm right-hand\ side})
  =\widetilde{S}(u_1+\xi_1+\xi_2 \tau-u_2)
  \widetilde{A}(u_1+\xi_1+\xi_2 \tau+\eta )
  \otimes \widetilde{A}(u_2)
\nonumber\\
  &&=e^{2 \pi i \xi_2 \eta (\gamma-1/N)} e^{-i \pi \tau \xi_2^2 /N
  -2 \pi i \xi_2 (u_1+\eta) /N}
\nonumber\\
  && \times (J_{{\bf \xi}} \otimes 1 ) \widetilde{S}(u_1-u_2)
  (J^{-1}_{{\bf \xi}} \otimes 1 )
  (J_{{\bf \xi}} \otimes 1 ) \widetilde{A}(u_1+\eta)
  \otimes \widetilde{A}(u_2)
  \nonumber\\
  &&=e^{-i \pi \tau \xi_2^2 /N-2 \pi i \xi_2 u_1 /N}
  e^{ 2 \pi i \xi_2 \eta (\gamma-2/N)}
  (J_{{\bf \xi}} \otimes 1 ) \widetilde{S}(u_1-u_2)
  \widetilde{A}(u_1+\eta) \otimes \widetilde{A}(u_2)  .
\label{e4-35}
\end{eqnarray}
In order that the transformation is the same in the left-hand side and 
the right-hand side of the Zamolodchikov algebra, we have $\gamma=2/N$
in case 2).\\
By the similar calculation, we have shown that the Zamolodchikov 
algebra has the same form  under the transformation 
$u_2\rightarrow u_2+\zeta_1+\zeta_2 \tau, \ (\zeta_1,\zeta_2\in Z)$. \\
The numerical calculation by REDUCE suggests that another 
Zamolodchikov algebra Eq.(\ref{e4-33}) is satisfied for $N=2$  
but not satisfied for $N\ge3$. 

\setcounter{equation}{0}
\section{The Heisenberg algebra and the elliptic theta function}
In this section, we first review the well-known relation between 
the Heisenberg algebra and the elliptic theta function\cite{Mumford}.\\
If $N$ is the square of some integer, that is, $N=l^2, (l\in Z)$, 
we can connect the another state vector $\tilde{A}_k(u)$ , which 
appears in Zamolodchikov algebra of the Belavin model, with  
the representation of the Heisenberg algebra by the theta 
function with characteristics.\\
The Heisenberg algebra is constructed from two operators $S_b,\ T_a$,
which are defined by 
\begin{eqnarray}
  && S_b f(u)=f(u+b)  ,
\label{e5-1}\\
  && T_a f(u)=\exp( \pi i a^2 \tau + 2\pi i a u) f(u+a\tau) ,  
\label{e5-2}
\end{eqnarray}
for the function $f(u)$.
We define the theta function with characteristics  in the form
\begin{eqnarray}
  && \theta_{a,b}(u,\tau)=S_b T_a \theta(u,\tau)
   =\exp\{ \pi i a^2 \tau +2\pi i a (u+b)\}\theta(u+a\tau+b,\tau)
\nonumber\\
  &&=\sum_{n\in Z} \exp\{ \pi i (n+a)^2 \tau +2\pi i (n+a) (u+b)\}  .
\label{e5-3}
\end{eqnarray}
Then we have 
\begin{eqnarray}
  && S_{\frac{1}{l}} \sum_{n\in \frac{1}{l}Z} c_n 
     \exp(\pi i n^2 \tau +2 \pi i n u) 
     =\sum_{n\in \frac{1}{l}Z} c_n \exp(2 \pi i n/l)
      \exp(\pi i n^2 \tau +2 \pi i n u)  ,
\label{e5-4}\\
  && T_{\frac{1}{l}} \sum_{n\in \frac{1}{l}Z} c_n 
     \exp(\pi i n^2 \tau +2 \pi i n u) 
     =\sum_{n\in \frac{1}{l}Z} c_{n-\frac{1}{l}}
      \exp(\pi i n^2 \tau +2 \pi i n u) .
\label{e5-5}
\end{eqnarray}
The action of $\displaystyle{S_{\frac{1}{l}}}$is given by
\begin{eqnarray}
  && S_{\frac{1}{l}} \sum_{n\in l Z} \exp(\pi i n^2 \tau +2 \pi i n u) 
     =\sum_{n\in lZ} \exp(\pi i n^2 \tau +2 \pi i n u)  ,
 \nonumber\\
  && S_{\frac{1}{l}} \sum_{n\in lZ+\frac{1}{l}} 
     \exp(\pi i n^2 \tau +2 \pi i n u) 
     =e^{2 \pi i/l^2}\sum_{n\in lZ+\frac{1}{l}} 
       \exp(\pi i n^2 \tau +2 \pi i n u)    ,
 \nonumber\\
  && \cdots
\nonumber\\
  && S_{\frac{1}{l}} \sum_{n\in lZ+\frac{l^2-1}{l}} 
     \exp(\pi i n^2 \tau +2 \pi i n u) 
     =e^{2 \pi i(l^2-1)/l^2}\sum_{n\in lZ+\frac{l^2-1}{l}} 
       \exp(\pi i n^2 \tau +2 \pi i n u)  .
\label{e5-6}
\end{eqnarray}
The action of $\displaystyle{T_{\frac{1}{l}}}$is given by
\begin{eqnarray}
  && T_{\frac{1}{l}} \sum_{n\in l Z} \exp(\pi i n^2 \tau +2 \pi i n u) 
     =\sum_{n\in lZ+\frac{1}{l}} \exp(\pi i n^2 \tau +2 \pi i n u)  ,
 \nonumber\\
  && T_{\frac{1}{l}} \sum_{n\in lZ+\frac{1}{l}} 
     \exp(\pi i n^2 \tau +2 \pi i n u) 
     =\sum_{n\in lZ+\frac{2}{l}} 
       \exp(\pi i n^2 \tau +2 \pi i n u)  ,
 \nonumber\\
&& \cdots
\nonumber\\
  && T_{\frac{1}{l}} \sum_{n\in lZ+\frac{l^2-1}{l}} 
     \exp(\pi i n^2 \tau +2 \pi i n u) 
     =\sum_{n\in lZ} \exp(\pi i n^2 \tau +2 \pi i n u)  .
\label{e5-7}
\end{eqnarray}
Then we take the basis, which is closed under actions $S_{\frac{1}{l}}$
and $T_{\frac{1}{l}}$, in the form
\begin{eqnarray}
  && {\bf v}(u)
  =\left(   \begin{array} {c}  v_0(u) \\ v_1(u) \\ v_2(u) \\
    \cdots \\ v_{l^2-1}(u)  \end{array} \right)
  =\left(   \begin{array} {c}  
  \sum_{n\in lZ} \exp(\pi i n^2 \tau +2 \pi i n u) , \\ 
  \sum_{n\in lZ+\frac{1}{l}} \exp(\pi i n^2 \tau +2 \pi i n u) ,  \\ 
  \sum_{n\in lZ+\frac{2}{l}} \exp(\pi i n^2 \tau +2 \pi i n u) , \\
    \cdots \\ 
  \sum_{n\in lZ+\frac{l^2-1}{l}} \exp(\pi i n^2 \tau +2 \pi i n u) 
  \end{array} \right)  .
\label{e5-8}
\end{eqnarray} 
Here we use $Z$ and $X$, which is the $l^2 \times l^2$ cyclic 
representation of $SU(2)$, in the form 
\begin{eqnarray}
  &&Z=\left(   \begin{array} {ccccc}  
  1 &  &  &   &  \\
    & \omega &  &  & \\
  &   & \omega^2  &  & \\
  &   &  & \cdots & \\
  &  &  &  & \omega^{l^2-1} \\ 
  \end{array} \right), \quad
  X=\left(   \begin{array} {ccccc}  
  0 & 0 & \cdots  & 0  & 1 \\
  1  & 0 & \cdots & 0 & 0 \\
  0  & 1 & \cdots & 0 & 0 \\
  &   &  \cdots  &  &  \\
  0 & 0 & \cdots & 1 & 0 \\
  \end{array} \right),
\label{e5-9}\\
  &&{\rm with}\quad ZX=\omega XZ, \quad \omega=e^{2 \pi i /l^2}, \quad 
  \omega^{l^2}=1. 
\label{e5-10}
\end{eqnarray}
The Heisenberg algebra, acting on the basis ${\bf v}$, is expressed 
in the form
\begin{eqnarray}
  S_{\frac{1}{l}}\left(   \begin{array} {c}  v_0(u) \\ v_1(u) \\ v_2(u) \\
    \cdots \\ v_{l^2-1}(u)  \end{array} \right)
  =\left(   \begin{array} {ccccc}  
  1 &  &  &   &  \\
    & \omega &  &  & \\
  &   & \omega^2  &  & \\
  &   &  & \cdots & \\
  &  &  &  & \omega^{l^2-1} \\
  \end{array} \right)
  \left(   \begin{array} {c}  v_0(u) \\ v_1(u) \\ v_2(u) \\
    \cdots \\ v_{l^2-1}(u)  \end{array} \right)
  =Z \left(   \begin{array} {c}  v_0(u) \\ v_1(u) \\ v_2(u) \\
    \cdots \\ v_{l^2-1}(u)  \end{array} \right)   ,
\label{e5-11}\\
  T_{\frac{1}{l}}\left(   \begin{array} {c}  v_0(u) \\ v_1(u) \\ v_2(u) \\
    \cdots \\ v_{l^2-1}(u)  \end{array} \right)
  =\left(   \begin{array} {ccccc}  
  0 & 1 & 0 & \cdots  & 0 \\
  0  & 0 & 1 & \cdots & 0 \\
  &   &   & \cdots  & \\
  0 & 0  & 0 & \cdots & 1\\
  1 & 0 & 0 & \cdots & 0 \\
  \end{array} \right)
  \left(   \begin{array} {c}  v_0(u) \\ v_1(u) \\ v_2(u) \\
    \cdots \\ v_{l^2-1}(u)  \end{array} \right)
  =X^{-1}
  \left(   \begin{array} {c}  v_0(u) \\ v_1(u) \\ v_2(u) \\
    \cdots \\ v_{l^2-1}(u)  \end{array} \right)  .
\label{e5-12}
\end{eqnarray}
Noticing that $ S_{\frac{1}{l}},T_{\frac{1}{l}}$ as the operator on
the basis but not the matrix, we have 
\begin{eqnarray}
  S_{\frac{1}{l}} T_{\frac{1}{l}} {\bf v}(u)
  =S_{\frac{1}{l}} X^{-1}{\bf v}(u)=X^{-1} S_{\frac{1}{l}}{\bf v}(u)
  =X^{-1} Z {\bf v}(u) . 
\label{e5-13}
\end{eqnarray}
\begin{eqnarray}
  \omega T_{\frac{1}{l}} S_{\frac{1}{l}}{\bf v}(u)
  =\omega T_{\frac{1}{l}} Z {\bf v}(u)
  =\omega Z T_{\frac{1}{l}} {\bf v}(u)
  =\omega Z X^{-1} {\bf v}(u) .
\label{e5-14}
\end{eqnarray}
Using $ZX=\omega XZ$, we have the fundamental relation 
of the Heisenberg algebra
\begin{eqnarray}
  S_{\frac{1}{l}} T_{\frac{1}{l}}=\omega T_{\frac{1}{l}} S_{\frac{1}{l}} .
\label{e5-15}
\end{eqnarray}
The standard basis is Eq.(\ref{e5-8}) with $l^2$-dimension. 
This is expressed as the linear combination of 
$\theta_{\frac{a}{l}, \frac{b}{l}}(u,\tau)$ with $(a=0, 1, 2, \cdots, l-1)$
and $(b=0, 1, 2, \cdots, l^2-1)$ in the form 
\begin{eqnarray}
  &&\theta_{\frac{a}{l}, \frac{b}{l}}(u,\tau)
  =\omega^{ab}\sum_{n\in lZ+\frac{a}{l}} \exp(\pi i n^2 \tau +2 \pi i n u)  
\nonumber\\
  &&+\omega^{ab}\omega^{lb}\sum_{n\in lZ+\frac{a+l}{l}} 
  \exp(\pi i n^2 \tau +2 \pi i n u)
\nonumber\\
  &&\cdots
\nonumber\\
  && +\omega^{ab}\omega^{l(l-1)b}\sum_{n\in lZ+\frac{a+l(l-1)}{l}} 
  \exp(\pi i n^2 \tau +2 \pi i n u) 
\nonumber\\
  &&= \omega^{ba}v_a+\omega^{b(a+l)}v_{a+l}
  +\cdots+\omega^{b\left(a+l(l-1)\right)}v_{a+l(l-1)}  .
\label{e5-16}
\end{eqnarray}
Multiplying $\displaystyle{ \frac{1}{l^2}\sum^{l^2-1}_{b=0} \omega^{-bc}}$ 
in the above relation, we have
\begin{eqnarray}
  &&v_a(u)=\frac{1}{l^2} \sum^{l^2-1}_{b=0}\omega^{-ba} 
        \theta_{\frac{a}{l}, \frac{b}{l}}(u,\tau), \nonumber\\
  &&v_{a+l}(u)=\frac{1}{l^2} \sum^{l^2-1}_{b=0}\omega^{-b(a+l)} 
        \theta_{\frac{a}{l}, \frac{b}{l}}(u,\tau), \nonumber\\
  && \cdots \nonumber\\
  &&v_{a+l(l-l)}(u)=\frac{1}{l^2} \sum^{l^2-1}_{b=0}
    \omega^{-b\left(a+l(l2-l)\right)} 
        \theta_{\frac{a}{l}, \frac{b}{l}}(u,\tau) ,
\label{e5-17}\\
   &&(a=0,1, \cdots,l-1) ,
\nonumber
\end{eqnarray}
where we use 
$\displaystyle{ \frac{1}{l^2}\sum^{l^2-1}_{b=0} \omega^{-bc}
\theta_{\frac{a}{l},\frac{b}{l}}(u,\tau)
=\delta_{c,a} v_a+\delta_{c,a+l} v_{a+l}+\cdots
 +\delta_{c,a+l(l-1)} v_{a+l(l-1)} } $ .
In this way, we have the two kind of basis. The state  
$v_p(u),\ (p=0,1, \cdots, l^2-1)$ is the vector basis of the Heisenberg algebra
with dimension $l^2$. While theta function with characteristics  
$\theta_{\frac{a}{l},\frac{b}{l}}(u,\tau),\ (a,b=0,1, \cdots (l-1))$ is the matrix 
basis  with dimension $ l \times l$. We summarize the bases of the Heisenberg 
algebra in the form
\begin{eqnarray}
  &&\hskip -10mm \theta_{\frac{a}{l}, \frac{b}{l}}(u,\tau)
 =\sum^{l-1}_{c=0}\omega^{b(a+lc)} v_{a+lc}(u), (a=0,1,2,\cdots, l-1) , 
 (b=0,1,2,\cdots, l^2-1) 
\label{e5-18}\\
  &&\hskip -10mm v_p(u)=\frac{1}{l^2}\sum^{l^2-1}_{b=0} \omega^{-bp} 
  \theta_{\frac{p}{l}, \frac{b}{l}}(u,\tau), 
  \quad (p=0,1,2,\cdots, l^2-1).
\label{e5-19}
\end{eqnarray}
The action of the Heisenberg algebra on these bases are 
\begin{eqnarray}
   &&S_{\frac{1}{l}} v_p(u)=\omega^p v_p(u), \quad
   T_{\frac{1}{l}} v_p(u)= v_{p+1}(u)  ,
\label{e5-20}\\
  &&S_{\frac{1}{l}}\theta_{\frac{a}{l}, \frac{b}{l}}(u,\tau)
  = \theta_{\frac{a}{l}, \frac{b+1}{l}}(u,\tau) , \quad
  T_{\frac{1}{l}}\theta_{\frac{a}{l}, \frac{b}{l}}(u,\tau)
  =\omega^{-b} \theta_{\frac{a+1}{l}, \frac{b}{l}}(u,\tau)  .
\label{e5-21}
\end{eqnarray}

\subsection{A property of the vector $v_p(u;\tau)$}
Here we will show 
$\theta_{\frac{p}{l^2},0}(l u; l^2\tau)=v_p(u; \tau), (p=0,1,\cdots, l^2-1), (N=l^2)$.
Then we can connect the state vector $\widetilde{A}_0(u)$, which is the special
case $k=0$ of the state vector which appears in the Zamolodchikov algebra 
of the Belavin model, with the basis of the Heisenberg algebra 
represented by the theta function with characteristics.\\
Starting from $v_p(u;\tau)$, we have
\begin{eqnarray}
  &&v_p(u;\tau)=\frac{1}{l^2}\sum^{l^2-1}_{b=0} \omega^{-bp} 
  \theta_{\frac{p}{l}, \frac{b}{l}}(u,\tau)
  =\frac{1}{l^2}\sum^{l^2-1}_{b=0} e^{-2 \pi i bp/l^2}
  \sum_{n\in Z}e^{i \pi \tau (n+\frac{p}{l})^2
  +2 \pi i (n+\frac{p}{l})(u+\frac{b}{l})}
\nonumber\\
  &&= \frac{1}{l^2} \sum^{l^2-1}_{b=0} e^{-2 \pi i bp/l^2}
  \sum^{l-1}_{c=0} \sum_{m\in Z}
  e^{i \pi l^2 \tau (m+\frac{c}{l}+\frac{p}{l^2})^2 
  +2 \pi i (m+\frac{c}{l}+\frac{p}{l^2})(lu+b)}
\nonumber\\
  &&=\frac{1}{l^2}\sum^{l^2-1}_{b=0} e^{-2 \pi i bp/l^2}
  \sum^{l-1}_{c=0} e^{2 \pi i b(\frac{c}{l}+\frac{p}{l^2})} 
  \theta_{\frac{c}{l}+\frac{p}{l^2},0}(lu;l^2\tau)
\nonumber\\
  &&=\frac{1}{l^2}\sum^{l^2-1}_{b=0} 
  \sum^{l-1}_{c=0} e^{2 \pi i b c/l} 
  \theta_{\frac{c}{l}+\frac{p}{l^2},0}(lu;l^2\tau)
  =\sum^{l-1}_{c=0} \delta_{c,0}e^{2 \pi i b c/l} 
  \theta_{\frac{c}{l}+\frac{p}{l^2},0}(lu;l^2\tau)
\nonumber\\
  &&= \theta_{\frac{p}{l^2},0}(lu;l^2\tau), \quad (p=0,1,2,\cdots, l^2-1).
\label{e5-22}
\end{eqnarray}
If we notice that $N=l^2$, we have 
\begin{eqnarray}
\theta_{\frac{p}{N},0}(lu; N\tau)=v_p(u; \tau),\quad (p=0,1,\cdots, N-1),
\label{e5-23}
\end{eqnarray}
which means that if we redefine $u$ as $lu$ in the Boltzmann 
weight $S(u)$, the state vector $v_p(u,\tau)$ in the Zamolodchikov algebra 
in case 2) is nothing but the representation of the Heisenberg algebra by 
the theta function with characteristics.

\vspace{10mm}
\section{Summary and discussion}
In oreder to understand the fundamental mechanism why the 
various two-dimensional statistical integrable model,
we have considered the typical spin type models, the Ising model and
the chiral Potts model, and the typical vertex type model, 
the Belavin model.
These typical integrable models are parametrized by the elliptic 
function in many cases. In this paper, we then have examined the connection 
between the elliptic function and the integrability condition.
We have shown that the integrability comes from the cyclic $SU(2)$ symmetry
of the model, which comes from the elliptic parametrization
of the Boltzmann weight.\\
The connection between the Heisenberg algebra and the elliptic theta 
function is well-known\cite{Mumford}, where Mamford has shown that the 
theta function with characteristics is the matrix representation 
of the Heisenberg albegra. In this paper, we have found that the vector 
representation of the heisenberg algebra. In the $N=l^2$ case,
we further have shown 
that the vector representation another of the Heisenberg albegra is 
equal to the state vector of the another Zamolodchikov algebra
of the Belavin model. 

\vspace{15mm}

\end{document}